\documentclass[nopubldata]{lpor2014}

\graphicspath{{figs/}}
\usepackage{float}

\begin{document}
%
\titlefigure{abstract.png}

\abstract{We discuss the progress in integration of nanodiamonds with photonic devices for quantum optics applications. Experimental results in GaP, SiO$_2$ and SiC-nanodiamond platforms show that various regimes of light and matter interaction can be achieved by engineering color center systems through hybrid approaches. We present our recent results on the growth of color center-rich nanodiamond on prefabricated 3C-SiC microdisk resonators. These hybrid devices achieve up to five-fold enhancement of diamond color center light emission and can be employed for integrated quantum photonics.}


\title{Nanodiamond integration with photonic devices}

%
\titlerunning{}
\author{Marina Radulaski\inst{1,*}, Jingyuan Linda Zhang\inst{1,*}, Yan-Kai Tzeng\inst{2}, Konstantinos G. Lagoudakis\inst{1}, Hitoshi Ishiwata\inst{3,4,5}, Constantin Dory\inst{1}, Kevin A. Fischer\inst{1}, Yousif A. Kelaita\inst{1}, Shuo Sun\inst{1}, Peter C. Maurer\inst{2}, Kassem Alassaad\inst{6}, Gabriel Ferro\inst{6}, Zhi-Xun Shen\inst{3,4}, Nicholas A. Melosh\inst{3,4}, Steven Chu\inst{2,7}, Jelena Vu\v ckovi\'c\inst{1}}
%
\authorrunning{M. Radulaski, J. L. Zhang, \emph{et al.}}
%
\institute{%
E. L. Ginzton Laboratory, Stanford University, Stanford, California 94305, United States
\and
Department of Physics, Stanford University, Stanford, California 94305, United States
\and
Geballe Laboratory for Advanced Materials, Stanford University, Stanford, California 94305, United States
\and
Stanford Institute for Materials and Energy Sciences, SLAC National Accelerator Laboratory,
2575 Sand Hill Road, Menlo Park, California 94025, United States
\and
Academy of Co-Creative Environment, Energy, Education and Science, Tokyo Institute of
Technology, 2-12-1 Ookaya-ma, Meguro-ku, Tokyo 152-8552, Japan
\and
Laboratoire des Multimateriaux et Interfaces, UMR CNRS 5615, Université de Lyon, 43
Boulevard du 11 Novembre 1918, 69622 Villeurbanne Cedex, France
\and
Department of Molecular and Cellular Physiology, Stanford University, Stanford, California
94305, United States
}
%
\mail{\email{marina.radulaski@stanford.edu; ljzhang@stanford.edu}}
%
%
\maketitle

\section{Introduction}

The paradigm-shifting concept of quantum information processing has been gaining practical relevance through the demonstrations of metropolitan \cite{sasaki2011field, valivarthi2016quantum} and satellite \cite{ren2017ground, liao2017satellite} quantum communication links, as well as the Noisy Intermediate-Scale Quantum (NISQ)-era quantum computers \cite{preskill2018quantum}. The problem of evolving these early prototypes to widespread commercial systems is an active research topic. Similarly to the development of classical electronic devices, material considerations play a defining role for quantum technologies. In evaluation of versatile quantum material systems, diamond color centers stand out as excellent single photon sources \cite{kurtsiefer2000stable, wang2005single}, long coherence qubit systems \cite{maurer2012room} and sensitive nano-magnetometers \cite{hong2013nanoscale}. Long-standing studies of the nitrogen-vacancy (NV) center have been continued by the centrally symmetric lattice defects such as the silicon-vacancy (SiV)\cite{rogers2014multiple}, neutral silicon-vacancy \cite{rose2018observation}, germanium-vacancy (GeV)\cite{iwasaki2015germanium, bhaskar2017quantum}, tin-vacancy (SnV) \cite{iwasaki2017tin, trusheim2018transform}, and lead-related optical centers \cite{trusheim2018lead, ditalia2018single, tchernij2018photoluminescence}, as well as by the emerging color centers whose origin is related to chromium defects and nickel-nitrogen complex \cite{aharonovich2010photophysics,aharonovich2010chromium,gaebel2004stable}.

Integration of diamond color centers with photonic devices provides a platform for quantum optical technologies. Nanophotonic devices can enhance the single photon emission rate \cite{zhang2018strongly} and quantum state readout sensitivity \cite{jensen2014cavity}. However, the development of these systems has started only in the past decade. While admirable progress in nanofabrication of bulk diamond has been made \cite{burek2012free, sipahigil2016integrated, wan2018two}, a scalable platform capable of hosting a variety of device geometries has not yet been reached. For example, the laborious diamond thinning into films \cite{li2015coherent} results in non-uniform thicknesses which significantly limits the area that can be exploited for photonics.  Innovative undercutting approaches \cite{burek2012free} have improved scalability but have, so far, provided only linearly and circularly shaped devices.

Hybrid approaches, where diamond color centers are interfaced with photonic devices implemented in a different substrate, are resulting in promising quantum platforms. Among them, nanodiamonds provide significant versatility in types of integration due to their small size and ease of production. In this review, we discuss the integration of nanodiamonds with photonic devices for quantum optics applications with an accent on our recent results in hybrid diamond-silicon carbide platform. We first introduce the physics of emitter-cavity systems, followed by the nanodiamond synthesis methods and their experimental integration with GaP, SiO$_2$ and SiC photonic platforms. For reviews on diamond spins, diamond plasmonics, broader photonics and single photon emission we refer the reader to references \cite{awschalom2018quantum, benson2011assembly, schroder2016quantum, aharonovich2016solid}.

\section{The physics of color centers embedded in photonic cavities}

The integration of quantum emitters with photonic cavities has been studied in a variety of systems involving atoms \cite{berman1994cavity}, quantum dots \cite{vuvckovic2003photonic}, and color centers \cite{riedrich2014deterministic}. The confinement of the optical field to small mode volumes results in enhanced light-matter interaction, advantageous for the development of fast and low-power communication devices. Here, we will discuss three regimes of Cavity Quantum Electrodynamics (CQED) relevant for diamond color center integration with cavities: Purcell enhancement, single emitter strong coupling and multi-emitter strong coupling.

\subsection{Color center emission}

The  emission spectrum of a color center consists of two sections, the zero-phonon line (ZPL) and the phonon sideband. The Lorenzian shaped ZPL represents a direct optical transition from the lowest vibrational excited to the lowest vibrational ground energy level. For the color centers with multiple ground or excited states, as in the case with the negatively charged silicon-vacancy, germanium-vacancy and tin-vacancy centers, the spectrum contains multiple ZPLs. The broad phonon sideband contains lower energy photons resulting from phonon-assisted optical transitions. The ratio of light emitted into the ZPL is called the Debye-Waller (DW) factor $\eta_{DW}$, and most of the nanophotonic applications benefit with higher values of this parameter. For NV center, the DW factor is 3-5\% \cite{wolters2010enhancement}, while inversion-symmetric color centers exhibit higher values. The negatively charged SiV center emits 70\% of radiation into its four ZPLs, with a maximum of 24\% for an individual transition \cite{zhang2018strongly}, while its neutral version SiV$^0$ emits 90\% into its only ZPL \cite{rose2018observation}. Negatively charged GeV and SnV centers exhibit a DW factors of 60\% \cite{siyushev2017optical} and 40\% \cite{thiering2018ab} across four ZPLs, respectively.

\subsection{Emitter-cavity systems}
A resonant emitter-cavity system at frequency $\omega$ is characterized by the number of emitters $N$, their single photon emission rate $\gamma$, the cavity energy loss rate $\kappa = \frac{\omega}{Q}$ inversely proportional to the cavity quality factor, the mode volume of the cavity $V$, and the individual emitter-cavity coupling $g \sim \sqrt{\frac{\eta_{DW}}{V}}\left|\frac{E(r)}{E_{max}}\right|$ \cite{radulaski2017nonclassical}. For solid state emitters that we discuss here, and assuming cryogenic temperatures, the emission rate is usually the smallest of the interaction rates in the system $\gamma \ll \kappa,g$. The light and matter interaction strength, expressed by $g$, depends on the electric field at the location of the emitter $E(r)$ relative to the maximum of the cavity field $E_{max}$. This implies that for the highest achievable coupling, the emitter should be positioned at the cavity field maximum, however, significant interaction can be achieved in hybrid approaches where the emitter couples to the evanescent tail of the cavity. Another aspect determining the coupling strength is the Debye-Waller factor $\eta_{DW}$. In the optimal case most of the color center emission would be through the ZPL. Therefore, some of the promising color centers for photonic applications include centrally symmetric color centers in diamond where the ZPLs take over 70\% of the emission.

\subsection{Purcell regime}

In the Purcell regime, the emission rate of the quantum emitter is enhanced. The presence of the resonant cavity alters the density of states in the system making the optical transition more probable and reducing the emission lifetime. Purcell effect occurs in small mode volume cavities where the emitter-cavity coupling is lower than the cavity loss rate, $g < \kappa/4$. The emission enhancement is characterized by the Purcell factor $F \sim \eta_{DW}\frac{Q}{V}$, which is highest for small lossless cavities with an emitter located at the field maximum. Additionally, Purcell enhancement of the ZPL emission results in the reduced emission of photons to the phonon sideband modes. This regime is used for fast and efficient generation of single photons in quantum networks, currently achieving up to 10-fold lifetime reduction in SiV centers in diamond nanocavities \cite{zhang2018strongly}, from nanosecond to hundreds of picosecond time-scales.

\subsection{Strong CQED coupling}

When the single emitter-cavity interaction is the dominant interaction, the system is said to be in the strong CQED coupling regime. This regime occurs when the photon is more likely to be exchanged between the cavity and the emitter, than with the environment, $g > \kappa/4$, $\gamma/4$, where $\kappa$ and $\gamma$ are the cavity energy decay rate and the dipole decay rate respectively. Thereby, the system enters a quantum hybrid state of light and matter, the so-called polariton state, featuring a superposition of a photon propagating in the cavity and being absorbed by the emitter. This interaction introduces anharmonicity in the ladder of energy states, providing routes for quantum light generation via photon blockade and photon-induced tunneling \cite{faraon2008coherent,radulaski2017nonclassical} applicable in integrated quantum communication and quantum metrology.

\subsection{Multi-emitter strong CQED coupling}

In a cavity incorporating $N$ emitters whose individual emitter-cavity coupling rate $g$ is not high enough to reach strong CQED coupling, the whole ensemble can potentially achieve multi-emitter strong CQED coupling with a collective coupling rate of $\sim g\sqrt{N}>\kappa/4$. The condition to exhibiting this effect, also known as the cavity protection, is that the inhomogeneous emission linewidth of the ensemble is comparable to the cavity linewidth $\kappa$. Compared to other solid state emitters, such as quantum dots whose inhomogeneous broadening exceeds cavity rates by orders of magnitude, diamond color centers have suitable properties for achieving  multi-emitter strong coupling. The proposed benefits of this regime include the novel forms of photon blockade effect, as detailed in the review \cite{radulaski2017nonclassical}.

\section{Nanodiamond synthesis and color center incorporation}

Nanodiamonds containing single photon emitters (SPEs) are a key component for quantum information processing with diamond color centers in the hybrid approach. Several methods can be used to create these nanodiamonds. The detonation method and high pressure high temperature (HPHT) growth produce NV center-containing nanodiamonds on a large scale, while the meteorites act as a scarce source for nanodiamonds containing SPEs. Small diamonds of the order of 2 nm found on meteorites have been shown to host SiV centers \cite{vlasov2014molecular}. Recently, a new method for luminescent diamond synthesis based on mixtures of hydrocarbon and fluorocarbon compounds without catalyst metals was developed\cite{davydov2014production}, producing high quality SiV and other near infrared color centers with almost lifetime-limited linewidths\cite{jantzen2016nanodiamonds, tran2017nanodiamonds}. Chemical vapor deposition (CVD) offers another promising alternative, which we will discuss in more detail. 

In the CVD approach, the diamonds can be grown on various substrates, and the dopant type and nanodiamond size can be independently controlled, which combine to provide higher engineerability. Early works have focused on 
controlled synthesis of high quality micro/nano-diamonds by microwave plasma CVD with tunable size, dispersion and levels of perfection of the nanodiamonds\cite{stacey2009controlled}, as well as exploring introducing various color centers as single photon emitters in the synthesized nanodiamonds \cite{rabeau2005fabrication, aharonovich2009two, beha2012diamond}.
For the negatively charged SiV in particular, microwave plasma CVD-synthesized nanodiamonds were some of the first resources for bright single photon sources \cite{neu2011fluorescence,neu2011single,neu2013low}, and have helped elucidate the electronic and optical structure of the color center \cite{muller2014optical}. The growth substrates include silicon, silicon carbide, silica, and iridium on yttria-stabilized zirconia (YSZ) on silicon. The nanodiamond growth typically starts with seeding with smaller de-agglomerated nanodiamonds from commercial sources \cite{neu2011single, chang2008mass, tisler2009fluorescence}, which are spin coated or drop casted onto the substrates, or with molecular diamond, also known as diamondoid \cite{tzeng2017vertical, zhang2015hybrid}, chemically attached to the substrate. The seeded substrates are then subjected to a microwave plasma assisted CVD process with hydrogen methane mixture at high temperature. Negatively charged SiV centers are formed \emph{in situ} and incorporated into the nanodiamonds due to the presence of silicon atoms from the nearby silicon-containing substrate. Similar methods have been employed earlier to create fluorescent nanodiamonds containing single NV centers \cite{rabeau2007single}, which paved the way for many NV based quantum information processing explorations.

\section{Diamond integration with GaP photonics}

Gallium phosphide (GaP) presents an attractive opportunity to integrate diamond color centers into hybrid resonant photonic circuits. GaP features large refractive index, high single crystal quality, optical transparency at the transition frequencies of diamond color centers, and strong $\chi^2$ nonlinearity\cite{rivoire2009second,shambat2010tunable,rivoire2010sum,rivoire2011second}. Hybrid GaP nanophotonic devices on bulk single crystal diamond evanescently couple optical fields to the underlying NV centers \cite{fu2008coupling, barclay2009chip, thomas2014waveguide, gould2016efficient, schmidgall2018frequency}. This way GaP devices incorporate quantum emitters with long spin coherence time and narrow inhomogeneous and homogeneous optical linewidths. In the case of the GaP-on-diamond disk resonators, the NV center ZPL emission is first resonantly enhanced and then efficiently coupled to a waveguide. This method has demonstrated efficient emission of the ZPL photons into the guided mode after optical excitation, as a result of combined effects of Purcell enhancement and efficient coupling between the resonant devices and waveguides. 
This is important development because without nanostructuring only $\sim 3\%$ of the emission from NV center in bulk diamond is into its ZPL, which is further limited by the collection efficiency in an unstructured diamond due to the large refractive index. The emission energy of the optical-circuit-integrated NV centers could be tuned and partially stabilized by an electric field \cite{schmidgall2018frequency}. In combination with the optical Purcell enhancement of the emission this integration represents a step towards scalable on chip photonic circuits.

In contrast to the bulk approach, nanodiamond-GaP hybrid systems offer an alternative platform for assembling integrated diamond-based nanophotonic quantum technology. A demonstrated pick-and-place technique assembles systems from the bottom-up with nanoscale manipulation \cite{benson2011assembly}. As with any technology, there are trade-offs in comparison with the alternatives such as evanescent coupling to color centers in bulk diamond or structures fabricated from bulk diamond. First, the nanodiamond-GaP hybrid system allows for pre-selection of both the color center-containing-nanodiamond and the photonic devices, which significantly improves the yield and to a certain extent offsets the resources needed for the nano-manipulation and assembly. The color center could be selected for better coherence time and spectral stability, and the nanophotonic devices such as optical resonators could be selected for frequency matching to the color centers of choice. Second, assembly through nano-manipulation allows for placement of the preselected color center at the cavity field maximum in the lateral direction with nanometer precision, which compares favorably to state-of-the-art color center placement in bulk diamond via ion implantation \cite{sipahigil2016integrated} and GaP-bulk diamond systems \cite{gould2016efficient}. However, one of the main disadvantages of the hybrid material system is that the optical modes are often confined in the high refractive index dielectric material. Therefore, the modes only evanescently couple to diamond,
and the color center in diamond cannot reside at the electric field maximum. As detailed in Section 2.2, this limits the light-matter coupling strength. The improvements in the resonator quality factor and the utilization of the air-modes of the optical system are some of the directions to mitigate this impediment. The small diameter of the nanodiamonds hosting the optically active color center, typically of the order of 10 nm, alleviates the disadvantage that the emitter resides outside the GaP. Another issue suffered by the nanodiamond hybrid platform is the spectral diffusion of color centers such as NV in nanodiamonds, which should be addressed from both material science and color center optical physics perspectives.

 Purcell enhancement of spontaneous emission into the ZPL has been demonstrated in the hybrid nanodiamond-GaP system, where the ZPL of an NV center is coupled to a single mode of a photonic crystal cavity\cite{wolters2010enhancement}. This work highlights the advantage of the bottom-up assembly approach: the color center and cavities are both pre-selected, and then tuned into resonance with each other. 
The Purcell enhancement mitigates the limitations of the natural physical parameters of the NV center, such as the small coupling strength between the NV center and electromagnetic field, and the small Debye-Waller factor.

 \begin{figure}
    \centering
    \includegraphics*[width=0.8\linewidth]{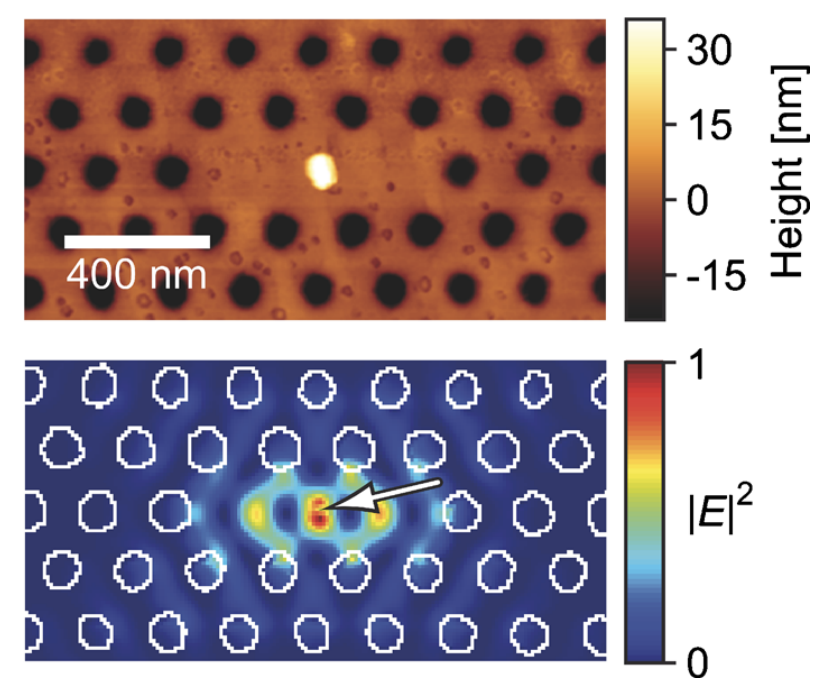}
    \caption{(a) AFM image of the GaP L3-cavity with a nanodiamond with 35 nm height placed near the center of the cavity. (b) FDTD simulated electric field profile of the fundamental mode. Figure reproduced with permission from \cite{wolters2010enhancement}. Copyright 2010 American Institute of Physics.}
    \label{fig:gap}
  \end{figure}

An L3 photonic crystal cavity is implemented in a 70 nm thick free standing GaP slab shown in Figure \ref{fig:gap}(a). The light is confined to a subwavelength mode volume of $\sim 0.75(\frac{\lambda}{n})^3$ with experimentally measured quality factors of $Q \sim 1000$ at  wavelengths near the NV center ZPL. Photo-oxidation of the GaP membrane is then used to fine-tune the cavity to the ZPL. 
Next, diamond suspension is spun onto a glass cover slip for the pre-characterization step where single-emitter nanodiamonds are identified using the pulsed second-order correlation measurement.
A suitable nanodiamond is then transferred to the GaP membrane using the pick-and-place technique\cite{schell2011scanning}. 
The nanodiamond dimensions (Figure \ref{fig:gap}(a)) imply that the NV center is within 35 nm from the GaP surface, which provides appreciable field strength at the NV center location even though the evanescent field decays from the surface. The light-matter interaction results in enhanced ZPL emission with an experimentally measured Purcell factor of $F_{exp}=12.1$, within an order of magnitude from the modeled value of $F_{model}=61$.

\section{Nanodiamond integration with silica photonics}

\subsection{Silica resonators}
Silica microdisk and microsphere optical resonators have been widely studied as high quality factor silicon photonic compatible devices for a variety of applications, including frequency combs\cite{kippenberg2011microresonator}, Brillouin lasers\cite{grudinin2009brillouin, morrison2017compact}, cavity optomechanical oscillators\cite{kippenberg2008cavity}, reference cavities\cite{matsko2007whispering}, and the study of cavity quantum electrodynamics \cite{aoki2006observation}. Ultrahigh-quality (UHQ) factor resonators have been extensively developed using conventional semiconductor processing methods on a silicon wafer. As in GaP nanophotonic resonators, silica optical resonators have been used to couple to color centers in nanodiamonds for quantum information processing and cavity quantum electrodynamic applications\cite{aoki2006observation}.

\begin{figure}
    \centering
    \includegraphics*[width=\linewidth]{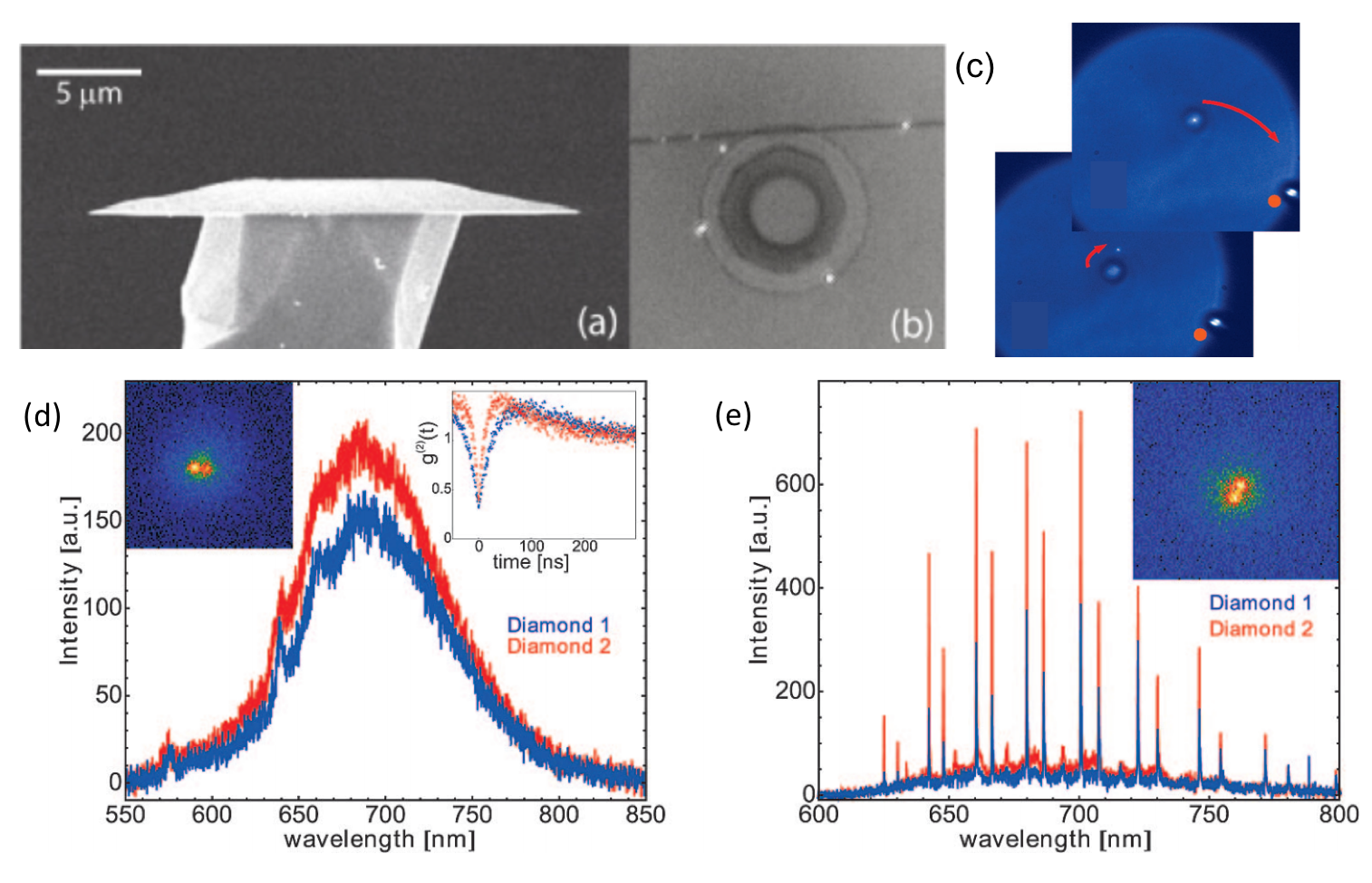}
    \caption{(a) SEM image of the sideview of a silica microdisk. (b) Optical image of the top view of the microdisk coupled to three diamond nanoparticles.  (c) The process of attaching a nanodiamond to the sphere. The sphere at the tip (round) is approached to the diamond (bright dot on substrate). After touching the nanodiamond, the diamond is picked up by the sphere. (d) Fluorescence spectra of two nanodiamonds before the transfer. Inset: second order correlation result $g^{(2)}(\tau)$ of the fluorescence from the two color centers. (e) Fluorescence spectra of the same nanodiamonds as in (d) attached to the sphere and individually excited by the focused laser, demonstrating coupling to the same set of WGMs.
   Figure (a-b) reproduced with permission from \cite{santori2010nanophotonics}. Copyright 2010 IOP publishing.
    Figure (c-e) reproduced with permission from \cite{schietinger2008one}. Copyright 2008 American Chemical Society.}
   \label{fig:silica1}
\end{figure}

In one of the early attempts \cite{santori2010nanophotonics}, diamond nanocrystals were attached to silica microresonators to achieve CQED effects. Silica microcavities 20 $\mu$m in diameter were fabricated, with initial quality factor of $Q=40000$. Nanodiamonds of the size of 70 nm were suspended in solution and deposited on the devices. After deposition of the nanoparticle and contact by tapered fiber, the low temperature quality factor reduced to $2000<Q<3000$. The Figure \ref{fig:silica1}(b) shows the resonant laser light scattered from three nanoparticles from this experiment. Confocal photoluminescence at low temperature collected from the top and through the fiber demonstrated the luminescence coupling to the whispering gallery modes (WGMs). By gas tuning the cavity modes, authors observed that the single NV center coupled to two different modes consecutively. However, no significant change in the spontaneous emission rate was observed, likely due to the large mode volume, limited quality factor, and spectral property of the emitters in nanodiamonds.

One of the main disadvantages of the hybrid-material system is that the emitter cannot be placed at the guided mode maximum, which can be largely overcome with continued improvements in resonator quality factor. It is worth noting that it is possible to retain the high quality factor of the resonator after transferring the nanodiamond onto the optical cavity. In the work by Barclay \emph{et al.} \cite{barclay2009coherent}, a fiber taper waveguide is used to both pick up and position the nanodiamonds onto the cavities, as well as a probing tool to measure the transmission of optical fields through the system. In this work, a quality factor of up to $8\times10^5$ was observed before transfer of the nanoparticle, and degraded to $1.7\times10^5$ after nanocrystal placement. Although other drawbacks, such as large mode volume and emitter quality, are yet to be resolved, this work demonstrates an important technique for deterministic placement while maintaining high $Q$ of the cavities. This research paves the way for future quantum optical experiments such as enhanced emission from the ZPL of color centers and cavity assisted control and readout of the quantum state. These assembly techniques combined with some state-of-the-art silica microdisk integrated silicon photonic circuits\cite{yang2018bridging} present a path towards on chip integration of nanodiamonds.

Using bottom-up assembly techniques, on-demand coupling beyond the single emitter to single cavity case is critical to building more complex quantum systems. In the work by Schietinger \emph{et al.} \cite{schietinger2008one}, two single NV centers in two nanodiamonds are coupled to the same set of high $Q$ whispering gallery modes of polystyrene sphere microresonators. Fluorescent nanodiamonds of the mean size of 25 nm and polystyrene microspheres with diameters around 5 $\mu$m were dispersed onto a substrate. The photoluminescence spectra and the second order correlation of the color center luminescence in the nanodiamonds are pre-characterized to confirm the single emitter nature before the assembly (Figure \ref{fig:silica1}(d)). Then, near-field scanning optical microscopy (NSOM) tips are used to pick up the microspheres and to bring them close to the pre-selected nanodiamond with a single NV center. As the microsphere comes into contact with the nanodiamond and moves across the substrate, the nanodiamond can be attached to the sphere, as shown in Figure \ref{fig:silica1}(c). The same procedure could be repeated to attach a second pre-characterized nanodiamond to the same microsphere. Since both nanodiamonds can be excited individually, it can be verified that both color centers in the two nanodiamonds couple to the same set of whispering gallery modes of the microsphere, as shown in Figure \ref{fig:silica1}(e). This technique could be extended to adding microsphere resonators to the coupled system to form photonic molecules, and moreover, opens the door to assembling more complex systems with arbitrary numbers of emitters and resonators. This demonstrated method combined with the strong coupling system parameters described below could provide a route towards multi-emitter CQED systems.

 \begin{figure}
    \centering
    \includegraphics*[width=\linewidth]{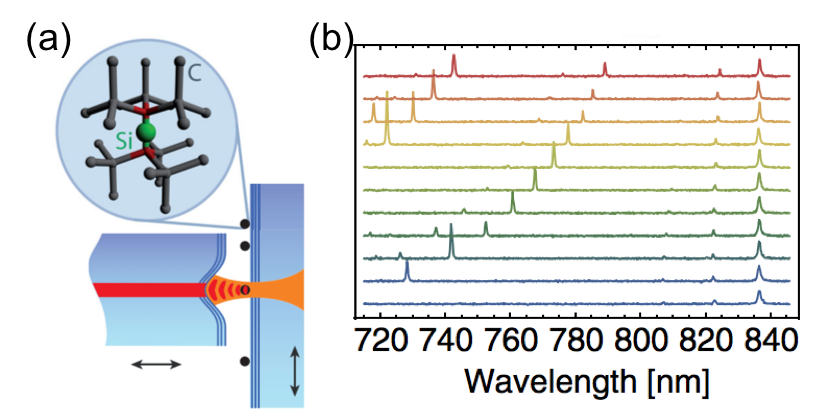}
    \caption{(a) Schematic of a fiber-based scanning cavity microscope cavity probing a SiV center in diamond. (b) Cavity transmission spectra for different cavity lengths. 
    Figure reproduced with permission from \cite{benedikter2017cavity}. Copyright 2017 American Physical Society.}
    \label{fig:silica3}
\end{figure}

Enhanced light-matter interaction between a diamond NV centers and a microresonator was achieved with a diamond nanocrystal coupled to a silica microsphere \cite{park2006cavity}. Silica microspheres feature WGMs with high quality factors $Q > 10^8$. This experiment was enabled by several technological advances: first, the ability to free-space evanescently couple to the WGMs using nearly spherical but non-axisymmetric microspheres, and more fundamentally the ability to fabricate these high $Q$ resonators by heating the fiber tip and merging microspheres using CO$_2$ lasers; second, the ability to conduct low temperature transmission measurements; and thirdly, the availability of stable, narrow, tunable laser sources at NV ZPL wavelength which allows for measurements of the transmission with high resolution and signal to noise ratio. The high $Q$ WGMs are excited by focusing a laser beam at $45^{\circ}$ from either the major or minor axis of the equator just outside the sphere surface. The diameters of the spheres are $\sim35$ $\mu$m with $Q$ approaching $10^8$ and FSR around the inhomogeneous distribution of the NV centers. The transmission spectrum of the resonator is obtained through a resonant scattering geometry. Nanodiamonds with average size 75 nm were irradiated with electrons followed by high temperature anneals to generate NV centers, and then they are dispersed onto the microsphere using solution deposition. The large number of small nanocrystals degrade the cavity linewidths to 20-80 MHz, which is comparable to the linewidth of individual NV centers. Next, the frequency of the quantum emitter and cavity mode are tuned to match each other using temperature tuning and coherent effects are observed. This work is an important demonstration of robust optical stability in the strong coupling CQED regime, and combined with other assembly techniques described in the previous sections, can be used to achieve scalable multi-cavity, multi-emitter platforms.

In addition to static cavity geometries such as microdisks and microspheres, a fiber-based microscopic and spectroscopic technique called scanning cavity microscope  \cite{mader2015scanning} can be applied to color centers such as NV and SiV in diamond to achieve high efficiency, high brightness, and high purity single photon sources\cite{albrecht2013coupling, kaupp2016purcell, benedikter2017cavity, kaupp2013scaling}. In these systems, a Fabry-Perot (FP) cavity is formed between a laser-machined and mirror-coated optical fiber and a coated planar mirror which also serves as the sample holder, as shown in Figure \ref{fig:silica3}(a). The coatings ensure that 90\% of the light is emitted into the fiber, which serves as the collection channel. The transverse movement of the sample mirror is used for spatial imaging of the features, and axial movement of the fiber tunes the resonance frequencies of the FP cavity, as shown in Figure \ref{fig:silica3}(b). The high quality factor of $1.9\times10^4$ combined with a mode volume of $3.4(\frac{\lambda}{n})^3$, results in a Purcell factor of $F = 9.2$ and 90\% collection efficiency at room temperature. This geometry is particularly desirable for fiber-coupled, high rate (4.1 MHz) single photon source in the ambient environment, suitable for applications such as quantum key distribution and all-optical quantum computing.

\subsection{Silica waveguides}

An alternative approach to nanodiamond color center integration is through direct coupling to an optical fiber. Various transfer techniques can be utilized to assemble alignment free, micrometer-scale single photon sources, by placing the NV nanodiamonds at the photonic crystal fiber end facets \cite{schroder2010fiber} or nanofiber-waist of a tapered optical fiber \cite{liebermeister2014tapered, vorobyov2016coupling}. Up to 689 kcps single
photon emission rate at saturation has been observed from a single NV center \cite{schroder2012nanodiamond}.
By embedding nanodiamonds containing NV centers into tellurite soft glass, color centers can be integrated close to the field maximum of the guided mode rather than though conventional evanescent coupling \cite{henderson2011diamond}.

\section{Nanodiamond integration with SiC photonics}

Integration of diamond with silicon carbide brings about new photonics opportunities due to the similarity of these two substrates. Like in GaP and SiO$_2$-diamond hybrid platforms, SiC also has a wide band gap and can support the propagation of visible and NIR wavelengths. Compared to the SiO$_2$-diamond hybrid platforms, devices can be made smaller and more scalable. In contrast to III-V substrates, SiC is made of group-IV elements, like diamond and silicon, and therefore lends itself to CMOS compatible processing which is reflected in 3C-SiC film growth on silicon wafers \cite{soueidan2006vapor}. Moreover, the refraction indices of SiC ($n_{SiC} = 2.6$) and diamond ($n_D = 2.4$) are similar in value which enables propagation of light field through both media, avoiding the localization of light into one substrate typical of most other hybrid platforms. Finally, silicon carbide is a $\chi^{(2)}$-nonlinear material that could potentially be exploited for efficient frequency conversion between telecom and diamond color center wavelengths.

 \begin{figure}
    \centering
    \includegraphics*[width=0.8\linewidth]{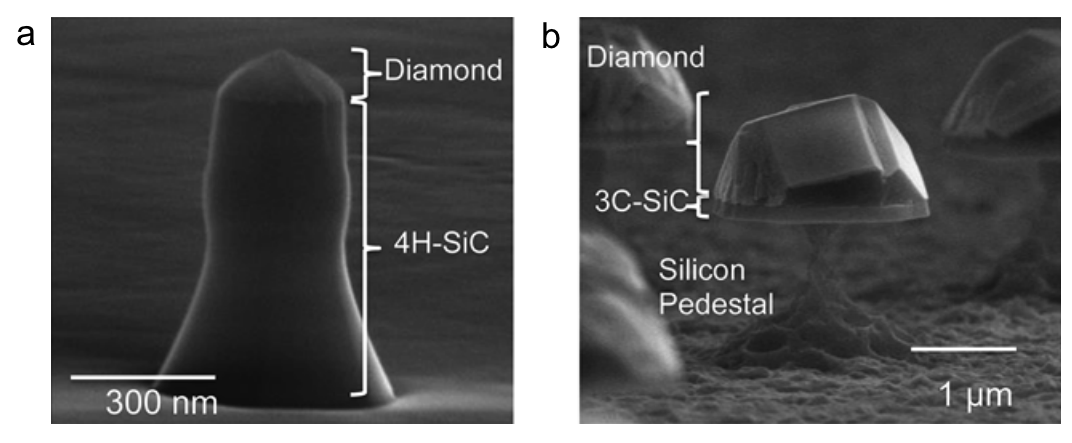}
    \caption{Nonresonant hybrid SiC-nanodiamond structures containing SiV color centers: (a) nanopillar and (b) nanodome. Figure adapted with permission from \cite{zhang2015hybrid}. Copyright 2015 American Chemical Society.}
    \label{fig:diamondoid}
  \end{figure}
  
Diamondoid-seeded CVD growth of nanodiamonds on 4H- and 3C-SiC substrates \cite{zhang2015hybrid} demonstrated that these crystalline nanoparticles can host color centers active up to room temperatures. Moreover, due to the differences in chemical reactivity of diamond and SiC, nanodiamonds were also utilized as a hard etching mask for silicon carbide. Resulting devices, shown in Figure \ref{fig:diamondoid}, include transparent nanopillars with diamond color center tips, as well as suspended hybrid nanodiamond-SiC structures that could potentially support whispering gallery mode resonances.

\begin{figure}
    \centering
    \includegraphics*[width=0.8\linewidth]{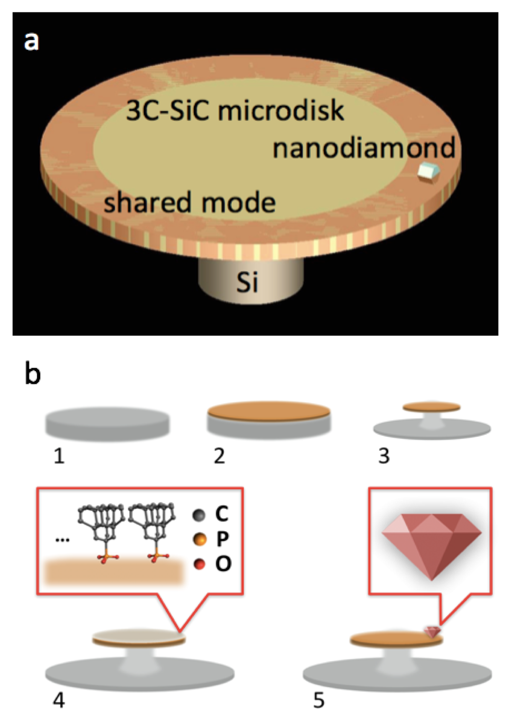}
    \caption{(a) Model of the hybrid silicon carbide-nanodiamond microresonator with a shared whispering gallery mode. (b) Growth and fabrication process for generating arrays of hybrid microresonators resulting in a preferential positioning of color center-rich nanodiamond at the periphery of silicon carbide microdisks on a silicon wafer.}
    \label{fig:SiC1}
  \end{figure}

Our recent work expands this approach to achieve Purcell enhancement of diamond color center emission. We develop a scalable hybrid photonics platform which integrates nanodiamonds with 3C-SiC microdisk resonators fabricated on a silicon wafer (Fig. \ref{fig:SiC1}). The nanodiamonds host SiV– and Cr-centers whose emission couples to high quality factor WGMs. The diamondoid-seeded CVD technique results in high yield and preferential positioning of color centers relative to the resonant mode. Up to 60\% of microresonators host nanodiamonds and in over 80\% of instances the nanodiamonds are located at the outer edge of the disks. The similarity in the refractive indices of diamond and silicon carbide facilitates penetration of a microdisk's evanescent field into the nanodiamond, resulting in an increased field overlap compared to what would be achievable in III-V substrates of similar design. As a proof of concept of a functioning hybrid group-IV photonics platform incorporating diamond color centers, we demonstrate up to five-fold enhancement of SiV$^–$ and Cr-center emission.

\subsection{Growth and fabrication}

The production of hybrid microresonators is based on the growth of group-IV materials on a silicon wafer (Fig. \ref{fig:SiC1}(b)). First, a 200 nm thick 3C-SiC layer is grown heteroepitaxially on (001)-Si using a two-step CVD method \cite{soueidan2006vapor}. Microdisks with diameters varying from 1.8-2.4 $\mu$m are lithographically defined and processed in plasma and gas phase etchers \cite{radulaski2014visible}, etching through silicon carbide and releasing most of the underlying silicon. Our goal is to deposit very small and pure color-center-rich nanodiamonds on top of the microdisks; therefore, we choose diamondoid as the best seeding candidate for the subsequent step. The microdisks are first exposed to oxygen plasma in order to generate an oxide layer. The seeding of [1(2,3)4] pentamantane diamondoids is covalently bonded to the oxidized surfaces of SiC microdisks via phosphonyl dichloride linkers for nanodiamond growth in a CVD chamber\cite{zhang2015hybrid}. Due to the charge concentration at the edges of the structure, the CVD plasma is denser around the ring-edge of the microdisk; as a result, the nanodiamonds that grow on microdisks are preferentially, in over 80\% of instances, positioned in the outer 30\% of the radius – in the vicinity of the whispering gallery modes.

The seeding diamondoids are relatively labile and prone to destruction and detachment from the substrate at high temperature and dense plasma during nucleation and etching. To overcome that, we modify the CVD growth to a two-step process. The surface of the microdisk was exposed to oxygen plasma for 5 minutes at 300 mTorr pressure and 100 W power. The diamondoids of [1(2,3)4] pentamantane as the seeding were covalently attached on the oxidized surfaces of SiC microdisk via phosphonyl dichloride bonding with SiO$_x$. The subsequent nanodiamond CVD growth process consists of two steps, one responsible for nanodiamond nucleation and the other determining the final nanodiamond size. The nucleation step is performed at 400 W power of Ar plasma, 350° C stage temperature and 23 mTorr pressure for 20 minutes. The flow of gases is: 5 sccm H$_2$, 10 sccm CH$_4$ and 90 sccm Ar. The duration of the second growth step varies between 15 and 45 minutes with 300 sccm H$_2$, 0.5 sccm [CH$_4$/SiH$_4$(1\%)] at 1300 W power of hydrogen plasma and 650° C stage temperature.

\begin{figure}
    \centering
    \includegraphics*[width=\linewidth]{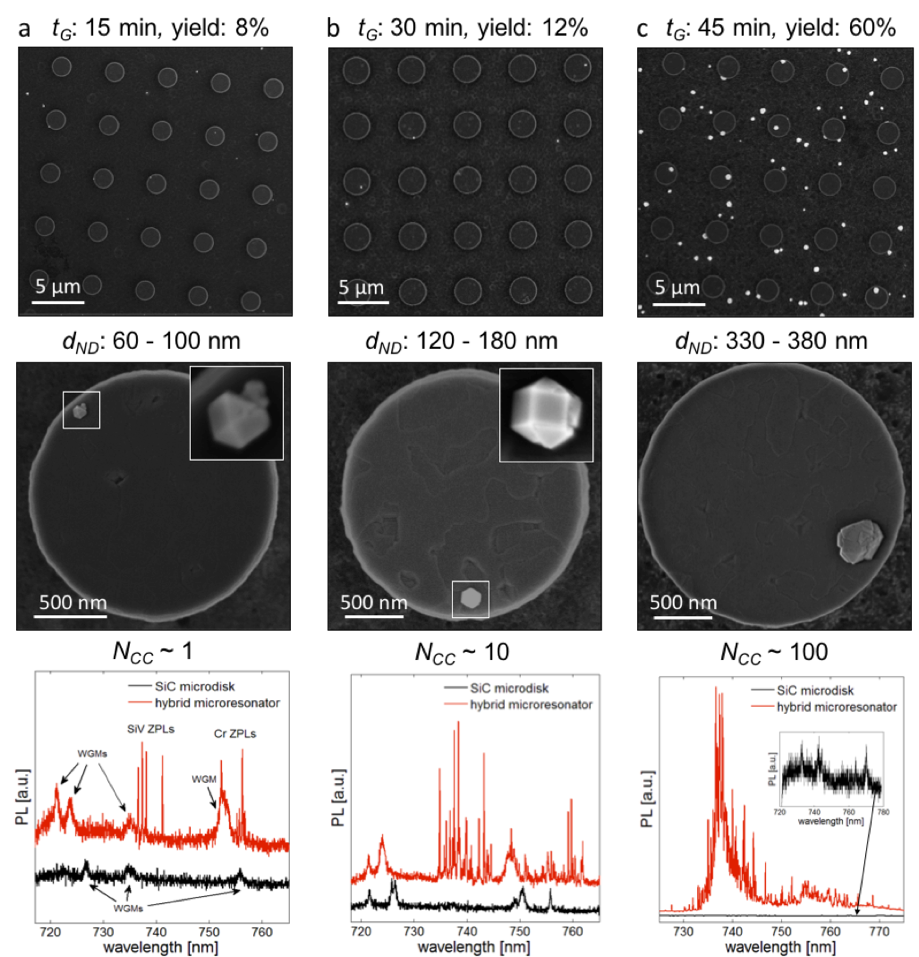}
    \caption{SEM images and photoluminescence signal at T = 10 K of three generations of hybrid silicon carbide-nanodiamond microdisk arrays. The variable growth time t$_G$ of a) 15 min, b) 30 min and c) 45 min determines the density of nanodiamonds on the chip, their diameter $d_{ND}$ and number of embedded color centers $N_CC$. Photoluminescence spectra comparison between resonators with and without diamond indicate the presence of whispering gallery modes with $\sim1$ nm linewidth.}
    \label{fig:SiC2}
  \end{figure}

SiV and Cr-related color centers are simultaneously generated as a result of in-situ doping with a controlled flow of silane  \cite{zhang2015hybrid} and by the uncontrolled doping with the residual chromium present in the chamber. The duration of the CVD growth defines the size and density of nanodiamonds on chip, as well as the number of color centers. The growth time $t_G$ is varied between 15 and 45 minutes. The smallest nanodiamonds are in the range 60-100 nm, and occur with 8\% yield, while the largest nanodiamonds are 330-380 nm in diameter and had 60\% yield, often resulting in multiple particles per microdisk (Fig. \ref{fig:SiC2}).

\subsection{Material characterization}

We estimate the order of magnitude of the number of color centers in the nanodiamonds by obtaining the photoluminescence spectra at cryogenic temperatures. Hereby, color center emission features $\sim100$ pm wide emission lines inhomogeneously broadened in the wavelength region 730-750 nm for SiV and 750-770 nm for Cr-centers. While recent results indicate that strained SiV centers can emit at wavelengths as long as 840 nm \cite{lindner2018strongly}, we conclude that in our case the emission at 750-770 nm originates dominantly from Cr-centers, since we observe two separate sets of peaks at a variety of temperatures and emerging as separate even in the smallest nanodiamonds (Fig. \ref{fig:SiC2}). To locate the sub-diffraction-limited diamonds we perform a 2D photoluminescence scan to identify the maxima of emission on hybrid microresonators and then spectrally resolve this signal. The spectra of the nanodiamonds indicate, depending on the diamond size, a few to several hundreds of color centers in an individual crystal (Fig. \ref{fig:SiC2}). The hybrid microresonator spectra also indicate $\sim1$ nm wide WGM resonances featuring a diameter-dependent free spectral range of 14-20 THz consistent with our previous SiC results \cite{radulaski2014visible}. Quality factors for hybrid microresonators are as high as 2700 for the case of 60-100 nm nanodiamonds, while the presence of larger, 330-380 nm, nanodiamonds reduces the quality factor to $Q < 1400$.

\subsection{Radiative enhancement}

\begin{figure}
    \centering
    \includegraphics*[width=\linewidth]{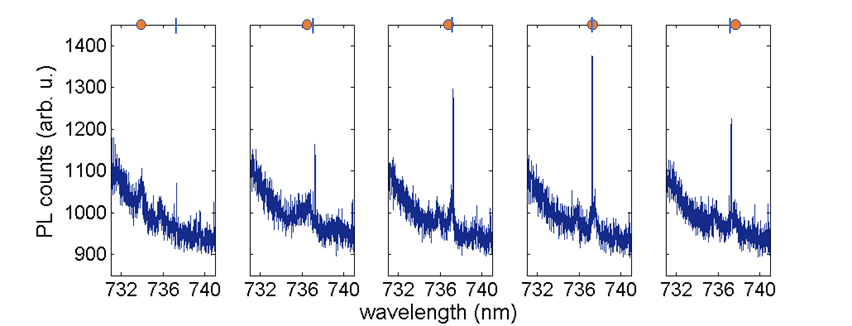}
    \caption{Photoluminescence of a hybrid microdisk with a 80 nm nanodiamond as the WGM is tuned across an SiV emission line. A three-fold enhancement in color center emission is observed for the resonant condition relative to the off-resonance condition. The red dot and the blue line mark the wavelengths of the WGM and the color center line, respectively.}
    \label{fig:SiC3}
  \end{figure}

Next, we investigate fluorescence enhancement in the color centers induced by the radiative coupling to a WGM. We first investigate at $T = 4$ K a microdisk with a small nanodiamond hosting a single color center at 737 nm and exhibiting a nearby resonance initially at 734 nm. Photoluminescence spectra are collected under 720 nm excitation using an NA = 0.9 objective, a confocal microscope and a spectrometer in front of a Si CCD. Using the argon gas condensation tuning technique, we continuously red-shift the resonance bringing it into and then out of the resonance with the color center by changing the effective index of refraction of the device. The photon count of the color center emission gets enhanced threefold at the resonant condition (Fig. \ref{fig:SiC3}) confirming that the coupling between a WGM and the color center is possible.

\begin{figure}
    \centering
    \includegraphics*[width=\linewidth]{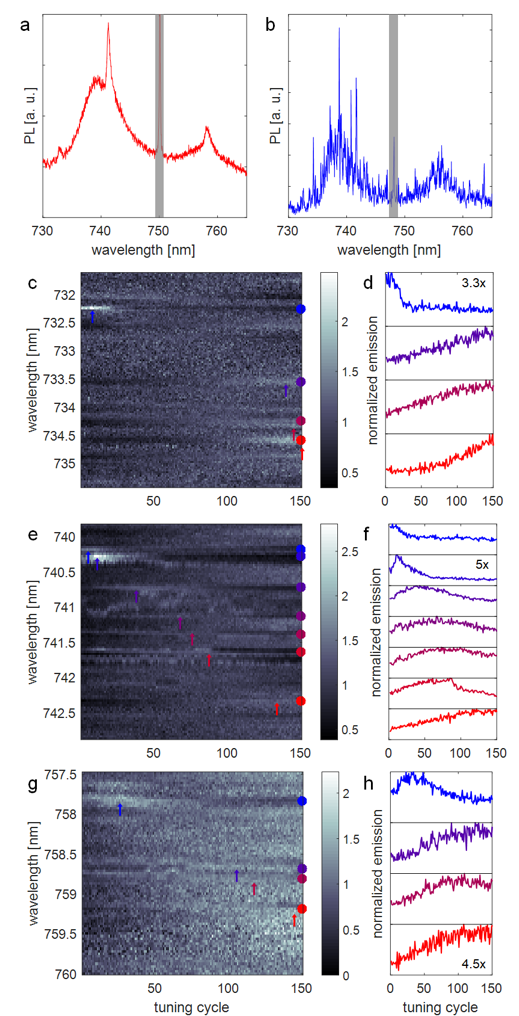}
    \caption{(a) Room temperature photoluminescence spectrum of a hybrid microdisk with a 350 nm nanodiamond supporting three resonances; the gray area covers a background induced peak. (b) T = 4 K spectrum of the same microresonator. (c-h) Resonantly-tuned T = 4 K photoluminescence spectra showing an increase in color center emission under the red-shifting resonance with plotted details of affected emitters. The wavelength intervals correspond to the three  whispering gallery mode resonances observed in (a); plot colors in (d),(f),(h) are color-coordinated with dots on the wavelength axes in (c),(e),(g); for each wavelength, the signal in (c),(e),(g) is normalized to the mean value obtained in 150 cycles (300 seconds) at that wavelength; y-axes of all subplots represent interval [0, 1], highest enhancement (max-to-min signal contrast) for each resonance is indicated in the relevant subplot.}
    \label{fig:SiC4}
  \end{figure}

\begin{figure}
    \centering
    \includegraphics*[width=\linewidth]{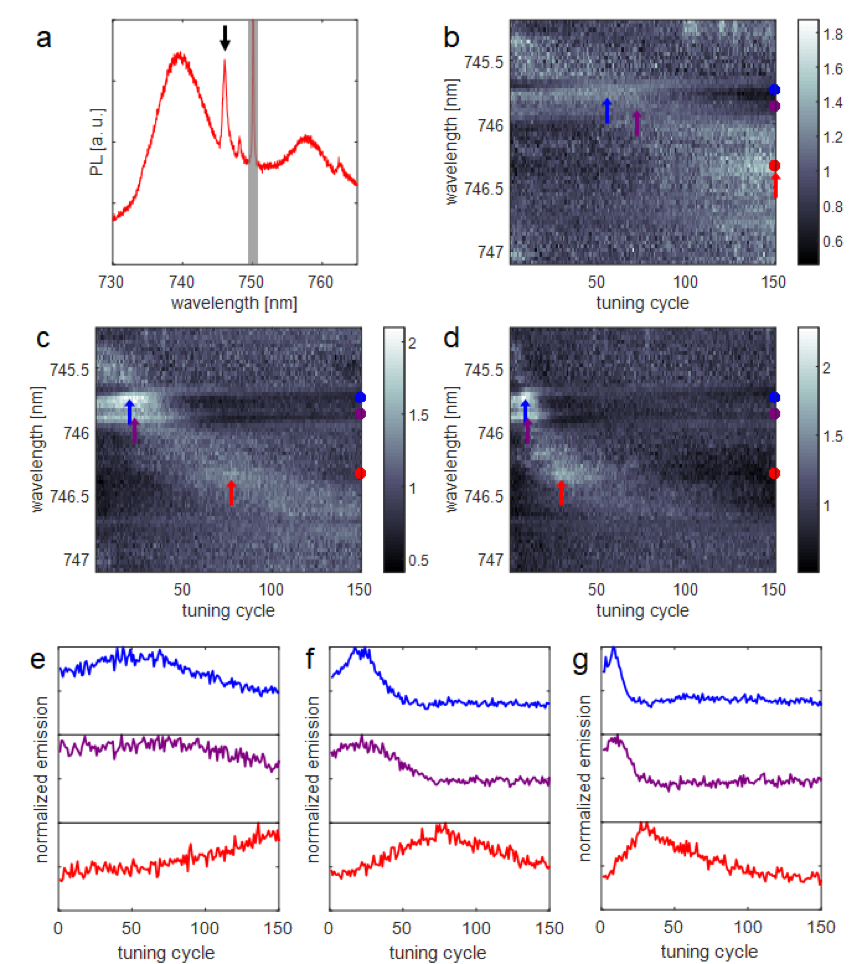}
    \caption{ (a) A room temperature spectrum of a hybrid microdisk with a 350 nm nanodiamond exhibiting three resonances; a black arrow indicates the analyzed WGM. Resonantly-tuned T = 4 K photoluminescence spectra showing an increase in color center emission under the red-shifting resonance with plotted details of affected emitters. The hybrid microdisks are argon-tuned by different flow rates (b) 0.1 sccm, (c) 0.3 sccm, (d) 0.5 sccm, and exhibit the same effect on a different time scale; for each wavelength, the signal is normalized to the mean value obtained in 150 cycles (300 seconds) at that wavelength; the arrows indicate peak enhancement of the color center emission. (e),(f),(g) Normalized color center emission from the color coded wavelengths indicated by dots in figures (b),(c),(d); y-axes of all subplots represent interval [0, 1].}
    \label{fig:SiC5}
  \end{figure}

Finally, we analyze a microresonator with a large nanodiamond and exhibiting three resonances observable at room temperature (Fig. \ref{fig:SiC4}(a)) that become obscured by color center zero-phonon lines at T = 4 K (Fig. \ref{fig:SiC4}(b)). We again tune the cavity resonance by gas condensation. It is worth noting that the tuning is not linear and the tuning rate is faster at the beginning of the process. The signal of some color center lines increases 2-5 times when in resonance with WGM. The observed difference in response among color centers is expected due to the different position and dipole orientation relative to the resonant electric field. This behavior is reproducible with different tuning rates set by the variable gas flow rate (Fig. \ref{fig:SiC5}). The possible causes of the observed signal enhancement are the Purcell enhancement and the resonant collection efficiency increase. To analyze the extent to which these effects can take place in our system, we turn to numerical modeling.

\subsection{Modeling and Discussion}

An in-house finite-difference time-domain (FDTD) code was used to model hybrid microdisks. We excite transverse electric (TE) and transverse magnetic (TM) modes with a 738 nm pulse with magnetic and electric field along the vertical axis, respectively, and analyze the electric field profile and the resonance quality factor in the cavity ring-down. The nanodiamonds are positioned at 85\% disk radius distance from the center, which has been determined as favorable in a separate analysis. The mode volume in all considered systems is $V\approx 5.5(\frac{\lambda}{n})^3$, while the resonant wavelength shift between the hybrid microdisks with different sizes of nanodiamonds is up to 1 nm.

\begin{figure}
    \centering
    \includegraphics*[width=\linewidth]{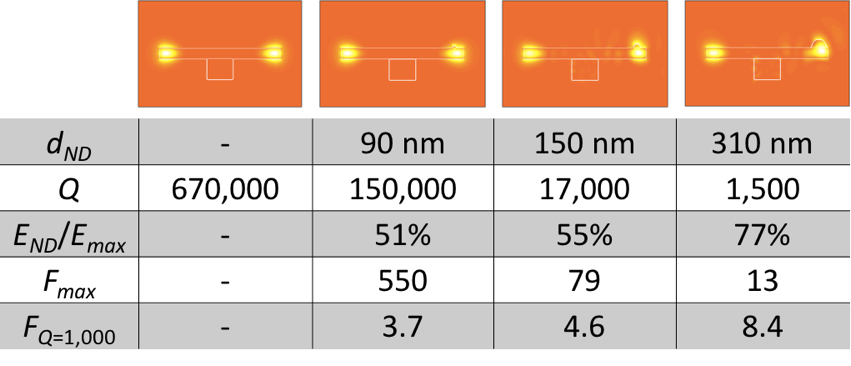}
    \caption{The analysis of differently-sized nanodiamond influence on a representative whispering gallery mode and its parameters in the hybrid microresonator, simulated by FDTD method.}
    \label{fig:SiC6}
  \end{figure}

\begin{figure}
    \centering
    \includegraphics*[width=\linewidth]{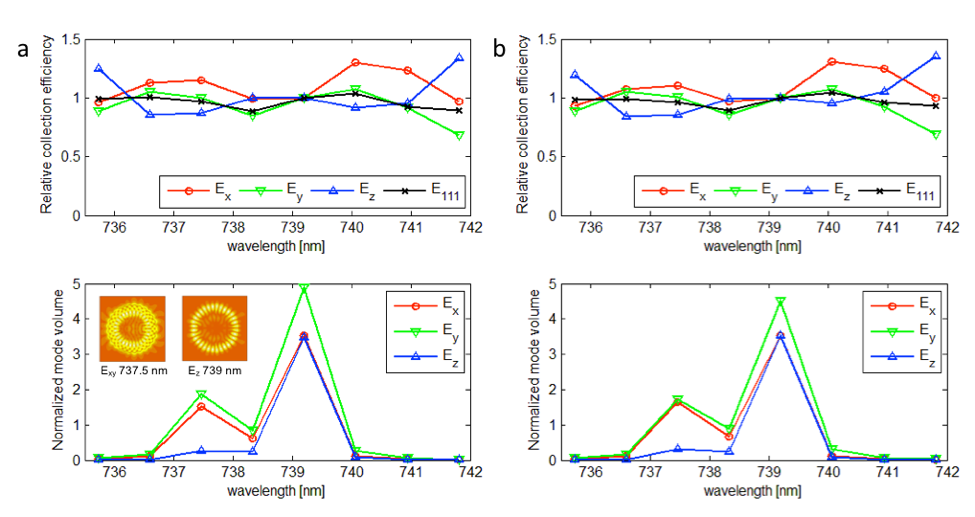}
    \caption{Wavelength-dependent collection efficiency and mode volume for (a) on-axis and (b) off-axis nanodiamond excitation in x-, y- and z- direction, as well as for 111-oriented dipole. Mode profile of the two emerging resonances is shown inset figure (a).}
    \label{fig:SiC7}
  \end{figure}

The FDTD simulations indicate that the microresonators' high quality factor, small mode volume and high electromagnetic field presence in the nanodiamond can facilitate color center emission Purcell enhancement of $F < 8$ (Fig. \ref{fig:SiC6}). The collection efficiency of light emitted from nanodiamond color centers with several different positions and three dipole orientations varies within a $\pm 35$\% margin between the on- and off-resonance condition (Fig. \ref{fig:SiC7}). It is worth noting that this model is not extensive due to the random positioning of nanodiamonds and the color centers within. Comparing the two modeling predictions to the experimentally observed 2- to 5-fold increase in the color center signal, we conclude that the Purcell enhancement is a likely cause of the color center signal increase. The demonstrated enhancement of color center emission is a proof of concept of a functional incorporation of diamond color centers into a group-IV photonics platform.

\section{Conclusions and Outlook}

Remarkable interdisciplinary progress has been made in the integration of nanodiamonds with photonic platforms for the development of color center-based quantum devices. The diversity of presented hybrid approaches, including GaP, SiO$_2$ and SiC substrates, show that nanodiamond light-matter interaction can be engineered for potential applications in quantum communication, computation and metrology. Additionally, due to biocompatibility, hybrid SiC and silica devices can be employed for extensions of nanodiamond \emph{in vivo} sensing research \cite{schirhagl2014nitrogen, childress2014atom}.

Studies of nanodiamond hybrid photonics have provided access to new device geometries relative to the bulk approaches. New nanodiamond synthesis methods promise to overcome strain and spectral diffusion issues and reach the ability for indistinguishable photon generation. Compared to other solid state quantum emitters, such as quantum dots, color centers provide new opportunities in scalability due to higher ensemble homogeneity. If light-matter interaction in diamond is increased to comparable intensities to those achieved in quantum dot-cavity systems, color center systems would not only reach the strong coupling of individual emitters, but also the collective strong coupling. Hereby, the color centers benefit from the ability to enter the cavity protection regime granted by the significantly smaller inhomogeneous broadening. Such regimes can be reached by advanced device design and fabrication, and impact photonics-based quantum simulation, distributed quantum communication and computation.


\begin{acknowledgement}
This work is supported by the National Science Foundation DMR Grant Numbers 1406028 and 1503759, the Air Force Office of Scientific Research under award number FA9550-17-1-0002, Army Research Office (ARO) Grant Number W911NF1310309, and the Department of Energy, Laboratory Directed Research and Development program at SLAC National Accelerator Laboratory, under contract DE-AC02-76SF00515.
\end{acknowledgement}

\begin{biographies}
  \authorbox{cvradulaski}{Marina Radulaski}{is a Nano- and Quantum Science and Engineering Postdoctoral Fellow at Stanford University, and an incoming Assistant Professor of Electrical and Computer Engineering at the University of California, Davis. She obtained a Ph.D. in Applied Physics from Stanford University in 2017 and was selected among the Rising Stars in EECS in 2017, Stanford Graduate Fellows 2012-2014, and Scientific American’s \emph{30 Under 30 Up-And-Coming Physicists} in 2012.}
  \authorbox{cvzhang}{Jingyuan Linda Zhang}{is a Ph.D. student in Applied Physics at Stanford University. She obtained her bachelor's degree in physics from Princeton University, and was supported by Stanford Graduate Fellowship 2013-2015. }
  \authorbox{cvtzeng}{Yan-Kai Tzeng}{is a postdoctoral fellow in Department of Physics at Stanford University. He obtained his Ph.D. in chemistry from National Taiwan University, Taiwan, M.S. in chemistry from National Taiwan Normal University, Taiwan, and B.S. in applied chemistry from National Chiayi University, Taiwan. }
  \authorbox{cvlagoudakis}{Konstantinos G. Lagoudakis}{is a Reader (Associate Professor) of Physics at the University of Strathclyde in Glasgow UK, leading research on hybrid quantum technologies. He obtained his Ph.D. in Physics from the Ecole Polytechnique Federale de Lausanne in Switzerland, M.S. in Optics and Photonics from Imperial College of Science and Technology in London, UK and B.S. in Physics from the National Kapodistrian University in Athens, Greece.}
  \authorbox{cvishiwata}{Hitoshi Ishiwata}{is a PRESTO researcher at the Japan Science and Technology Agency. He obtained a Ph.D. in Physics from Stanford University.}
  \authorbox{cvdory}{Constantin Dory}{is a Ph.D. student in Electrical Engineering at Stanford University. He obtained his BS and M.S. degrees from TU Munich in 2016, and is supported by the Andreas Bechtolsheim Stanford Graduate Fellowship. }
  \authorbox{cvfischer}{Kevin A. Fischer}{obtained his Ph.D. in Electrical Engineering from Stanford University. He obtained his bachelor's degree from Massachusetts Institute of Technology, and received Stanford Graduate Fellowship and the National Defense Science and Engineering Graduate Fellowship. }
  \authorbox{cvkelaita}{Yousif A. Kelaita}{obtained his Ph.D. in Electrical Engineering from Stanford University. He obtained his bachelor's degree from California Institute of Technology, and was supported by Stanford Graduate Fellowship and the National Defense Science and Engineering Graduate Fellowship. }
  \authorbox{cvsun}{Shuo Sun}{is a postdoctoral fellow at Stanford University. He received his M.S. (2015) and Ph.D. (2016) in the Department of Electrical and Computer Engineering from University of Maryland, College Park, and Bachelor of Science in 2011 with a specialization in optics from Zhejiang University in China. }
  \authorbox{cvmaurer}{Peter C. Maurer}{is an assistant professor in Molecular Engineering at University of Chicago. He received his postdoctoral training at Stanford University, Ph.D. in physics from Harvard University, and undergraduate degree in physics at the Swiss Federal Institute of Technology (ETH). }  
  \authorbox{cvalassaad}{Kassem Alassaad}{obtained a Ph.D. in Material Science in 2014 from the University Claude Bernard Lyon 1, France and a M.S. in 2011 from the Beirut Arab University, Lebanon.}  
  \authorbox{cvferro}{Gabriel Ferro}{is a senior Scientist at CNRS and the Adjoint Director of the Laboratoire des Multimat\'eriaux et Interfaces, University of Lyon, France. He obtained a Ph.D. in Material Science in 1997 from the University Claude Bernard Lyon 1, France. He was awarded the CNRS bronze medal in Chemical science on 2002. Since 2012, he is member of the steering committee of the European Conference on Silicon Carbide and related materials.} 
 
  \authorbox{cvshen}{Zhi-Xun Shen}{is the Paul Pigott Professor in Physical Sciences, and a senior fellow of the Precourt Institute for Energy, Stanford University. He received a B.S. degree from Fudan University in 1983, an M.S. degree from Rutgers University in 1985, and Ph.D. degree from Stanford University in 1989. Dr. Shen’s work has been recognized by important awards, including the Kamerlingh Onne Prize (international prize on superconductivity); the E.O. Lawrence Award (Department of Energy award); the Oliver E. Buckley Prize (condensed matter prize of the American Physical Society); Einstein Professorship (Chinese Academy of Sciences). Dr. Shen is a member of the National Academy of Sciences, and the American Academy of Arts and Sciences. }  
  \authorbox{cvmelosh}{Nicholas A. Melosh}{is an Associate Professor of Materials Science and Engineering at Stanford University and the Director of the Stanford Nanofabrication Facility. He received his Ph.D. in Materials Science and Engineering from the University of California, Santa Barbara in 2001, and a BS in Chemistry from Harvey Mudd College in 1996.
 }  
  \authorbox{cvchu}{Steven Chu}{is the William R. Kenan, Jr., Professor of Physics and Professor of Molecular and Cellular Physiology in the Medical School at Stanford University. He is the co-recipient of the 1997 Nobel Prize in Physics for his contributions to laser cooling and atom trapping, a member of the National Academy of Sciences, the American Philosophical Society, the American Academy of Arts and Sciences, the Academia Sinica, and is a foreign member of the Royal Society, the Royal Academy of Engineering, the Chinese Academy of Sciences, the Korean Academy of Sciences and Technology and the National Academy of Sciences, Belarus. He is the President Elect of the American Association for the Advancement of Science. He received an A.B. degree in mathematics and a B.S. degree in physics from the University of Rochester, and a Ph.D. in physics from the University of California, Berkeley, as well as 32 honorary degrees.
}  
  \authorbox{cvvuckovic}{Jelena Vu\v ckovi\'c}{is a Professor of Electrical Engineering and by courtesy of Applied Physics at Stanford University, where she leads the Nanoscale and Quantum Photonics Lab. She is also a faculty member of the Ginzton Lab, PULSE Institute, Simes Institute, and Bio-X at Stanford. She obtained a Ph.D. in Electrical Engineering from the California Institute of Technology (Caltech) in 2002. She holds a visiting position at the Institute for Physics of the Humboldt University in Berlin, Germany and is a Hans Fischer Senior Fellow at the Institute for Advanced Studies of the Technical University in Munich, Germany.}
\end{biographies}
%
\bibliographystyle{lpr}
\bibliography{references.bib}

\end{document}